\documentclass[twocolumn,showpacs,preprintnumbers,amsmath,amssymb,groupedaddress,superscriptaddress]{revtex4}
\usepackage{graphicx}% Include figure files
\usepackage{dcolumn}% Align table columns on decimal point
\usepackage{bm}% bold math

\begin{document}

%\doi{10.1080/0950034YYxxxxxxxx}
% \issn{1362-3044}
%\issnp{0950-0340} \jvol{00} \jnum{00} \jyear{2008} \jmonth{10 January}
%
%\markboth{E.\ E.\ Mikhailov \textit{et al.}}{Vacuum squeezing via polarization self-rotation  and excess noise in
%hot Rb vapors}
%
%\articletype{Research paper}

\title{Vacuum squeezing via polarization self-rotation  and excess noise in hot Rb vapors}

\author{Eugeniy E. Mikhailov}
\affiliation{The College of William $\&$ Mary, Williamsburg, Virginia, 23187}
\author{Arturo Lezama}
\affiliation{ Instituto de F\'{\i}sica, Facultad de Ingenier\'{\i}a, Casilla de correo 30, 11000, Montevideo,
Uruguay}
\author{Thomas W.\ Noel}
\affiliation{University of Wisconsin, 1150 University Avenue, Madison, WI 53706, USA}
\author{Irina Novikova}
\affiliation{The College of William $\&$ Mary, Williamsburg, Virginia, 23187}
%
%

%\author{
%Eugeniy E.\ Mikhailov$^{a}$, Arturo Lezama$^{b}$, Thomas W.\ Noel$^{c}$, and Irina
%Novikova$^{a}$$^{\ast}$\thanks{$^\ast$Corresponding author. E-mail: inovikova@physics.wm.edu \vspace{6pt}}\\
%$^{a}$ {\em{The College of William $\&$ Mary, Williamsburg, VA 23187, USA}};
% $^{b}$ {\em{ Instituto de F\'{\i}sica, Facultad de Ingenier\'{\i}a, Casilla de correo 30, 11000, Montevideo, Uruguay
% }}; $^{c}$ {\em{ University of Wisconsin, 1150 University Avenue, Madison, WI 53706, USA }}
% \\\vspace{6pt} \received{\today}}

%\maketitle %{for JMO style}

\begin{abstract}
We present experimental and theoretical analysis of quantum fluctuation in a vacuum field in the presence of
orthogonal linearly polarized pump field propagating through a Rb vapor cell. Previously reported theoretical
and experimental studies provided somewhat contradictory conclusions regarding the possibility to observe the
``squeezed vacuum'' -- the reduction of vacuum fluctuations below standard quantum limit -- in this system.
Here, using the $\mathrm{D}$1 transitions of Rb in a cell without buffer as as an example, we demonstrate that
vacuum squeezing is corrupted by incoherent processes (such as spontaneous emission, elastic scattering, etc.),
and its observation is only possible in a specific small region of the experimental parameter space. Numerical
simulations, in good agreement with the experiment, demonstrate that the two excited state hyperfine levels play
a crucial role in the squeezing and excess noise production. The significant influence of far-detuned atoms on
the field fluctuations at low noise frequencies imposes the explicit consideration of the full velocity
distribution of the atomic vapor.
\bigskip

%\begin{keywords}quantum fluctuations, squeezed states, polarization self-rotation, atomic noise
%\end{keywords}
\end{abstract}

\maketitle %{for APS style}

%\author{Eugeniy E. Mikhailov}
%\author{Irina Novikova}
%\affiliation{The College of William $\&$ Mary, Williamsburg, Virginia, 23187}
%
%
%\begin{abstract}
%\end{abstract}
%
%\pacs{
%    270.0270, % Quantum optics
%    270.6570, % Squeezed states
%    020.1670, % Coherent optical effect
%    020.6580, % Stark effect
%    140.0140  % Lasers and laser optics
%}
%%\definecolor{purple}{rgb}{0.6,0,1}
%%\preprint{\large \color{purple}{LIGO-P050005-00-R}}
\date{\today}
\maketitle

%\section{Introduction}

As a consequence of Quantum Mechanics, any optical
measurement that uses coherent light is fundamentally shot-noise limited. Not surprisingly, there
is a lot of interest in ways to reduce the measurement noise beyond this limit using the so-called
``squeezed'' states of light~\cite{bachor_guide_2004}.

The two orthogonal quadratures of the electromagnetic field are continuous variables that quantum-mechanically
correspond to the  non-commuting quadrature operators $S_\pm$~\cite{scullybook,bachor_guide_2004}. The
Heisenberg uncertainty principle dictates that the variances of the two quadratures obey: $\langle\Delta
S_+\rangle^2\langle\Delta S_-\rangle^2\ge \hbar^2/4$. For an optical field in a coherent state, the fluctuation
spectra of both quadratures are equal and correspond to the minimum allowed uncertainty. This minimum level of
fluctuations, that is the same for a vacuum field, is often called the ``standard quantum limit'' (SQL).
However, fluctuations in one of the quadratures may drop below the SQL as long as
fluctuations in the other quadrature increase to compensate  - this is the
characteristic property of squeezed light.
Squeezed light states are very fragile since optical losses and incoherent fluctuations added to the
squeezed quadrature reduce the amount of squeezing.

One of the most common ways to produce squeezed light relies on parametric down
conversion~\cite{bachor_guide_2004}. In this process one pump photon is converted to a pair of photons of lower
frequency. Due to conservation laws the frequency and momenta of the generated photons are strongly correlated.
While this method became a standard tool in nonlinear and quantum optics, it remains relatively complicated and
expensive. It is particularly challenging for generation of squeezed light in near-IR spectral region for atomic
physics applications, even though significant progress has been made in the recent
years~\cite{hetet_squeezed_at_D1_Rb_2007,akamatsu_ultraslow_2007, appel_sq_quantum_memory_Rb_2007,
honda_sq_storage_in_Rb_2007}.

\begin{figure}[h]
  \includegraphics[angle=0, width=1.0\columnwidth]{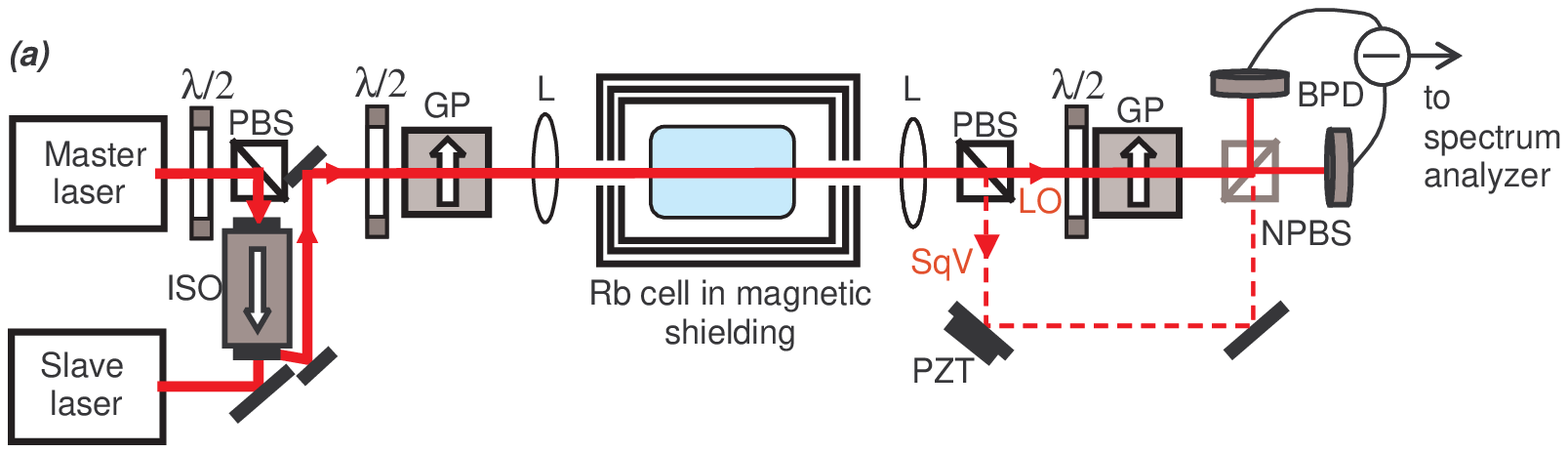}
 \includegraphics[angle=0, width=1.0\columnwidth]{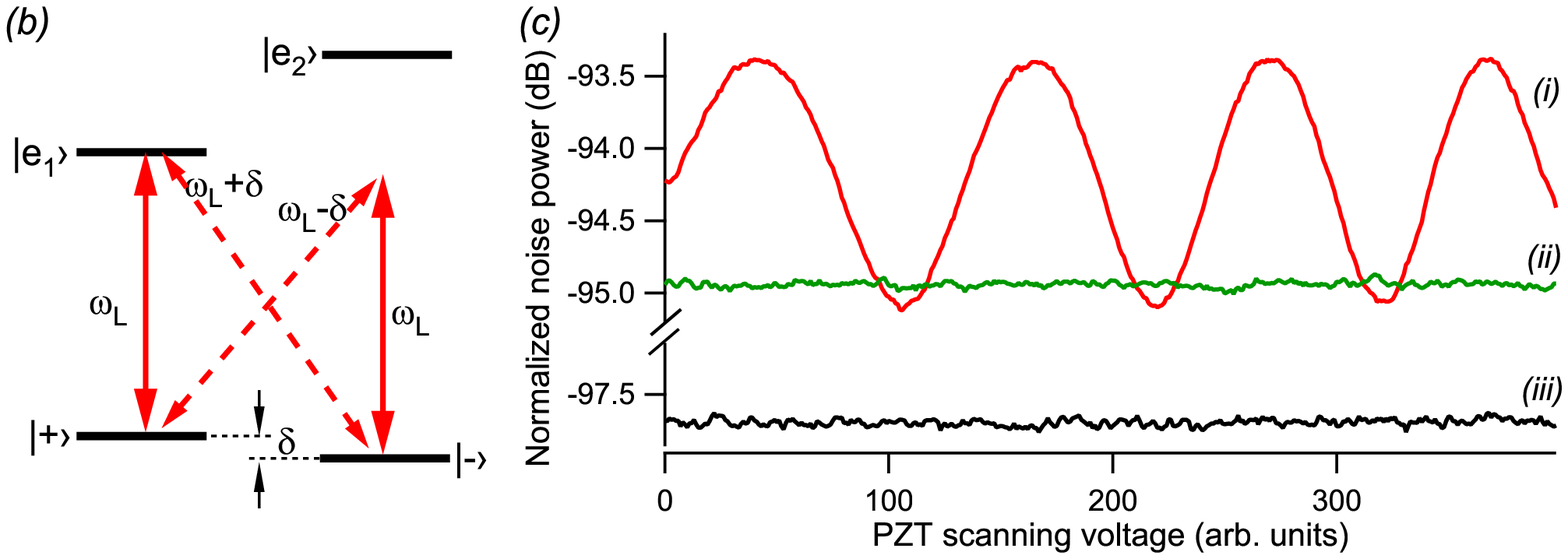}
  \caption{
    \emph{(a)} Schematic of the experimental setup (see the text for details).
    \emph{(b)} Simplified four-level scheme used to explain the  vacuum squeezing via PSR.
    \emph{(c)} Example of the detected quadrature noise $(i)$ dependence
    on the relative local oscillator phase (controlled by PZT scanning of the mirror); shot noise $(ii)$ and technical noise of the spectrum
    analyzer $(iii)$ are also shown. The data is taken at $1.36$~MHz sideband frequency (RBW=100kHz, VBW=30Hz). The laser
    frequency is near $F=2 \to F'=2$, and the temperature of the cell is $59^\circ$C
    ($N=2.3\cdot10^{11}~\mathrm{cm}^{-3}$).
    }
  \label{fig:apparatus}
\end{figure}

Recently, several research groups~\cite{
ries_experimental_2003,mikhailov2008ol,zibrovJETP2004,hsu_effect_2006,lezama_numerical_2008} explored another
method for generation of squeezed vacuum that is based on the rotation of elliptically polarized light as a
result of propagation through an isotropic optical material (when no external electric or magnetic field
present). Such polarization self-rotation (PSR) is a nonlinear optical effect that occurs in many optical media,
but it is particularly strong near resonant optical transitions in alkali metal atoms, where even a small
difference in resonant frequency due to different AC-Stark shifts of two circular components produces detectable
variation in their relative refractive
index~\cite{novikova_ac-stark_2000,rochester_self-rotation_2001,novikova_large_sr_squeezing_2002}.
Classically the PSR disappears for linearly polarized light; however, on the quantum level the effect
persists, changing the statistics of the vacuum field in the orthogonal polarization via cross-phase modulation
between two initially independent circular components. This results in quadrature  vacuum squeezing (VS) in the
orthogonal polarization at the medium output~\cite{matsko_vacuum_2002,novikova_large_sr_squeezing_2002}. The
advantages of this method are the relative simplicity of the experimental apparatus (Fig.~\ref{fig:apparatus}(a))
and the fact that the squeezed vacuum is automatically generated at frequencies corresponding to atomic
transitions.

The nonlinear light-atom interaction leading to VS via PSR corresponds (to the lowest order) to a four-photon
parametric process involving the absorption of two photons of the linearly polarized pump field and the
(correlated) emission of two photons symmetrically detuned from the pump field by the observed noise frequency
(noise sidebands) with orthogonal polarization. This picture is particularly useful to address the spectral
properties of the transmitted field noise in connection to the atomic level structure.
The simplest level scheme where such a process can take place is the four level configuration shown in Fig.
\ref{fig:apparatus}(b). In this system, a linearly polarized pump field couples two (Zeeman) ground states
$|\pm\rangle$ with two different excited states $|\mathrm{e}_{1,2}\rangle$. The other two transitions from
ground to excited states correspond to the orthogonal polarization. Because of the difference in coupling and
detuning of the two driven optical transitions, the energies of the ground states $|\pm\rangle$ can be slightly
different as a consequence of differential AC Stark shifts. The vacuum fluctuations, in the orthogonal
polarization, at a frequency corresponding to the ground state energy splitting,  are in two-photon resonance.
Under this condition, the parametric four-wave process can efficiently produce phase-dependent correlation
between the noise sidebands leading to the possibility of squeezing. The sideband bandwidth  for potential
vacuum squeezing is determined by the decoherence rate of the ground states which depends on the intensity and
frequency of the pump field and on the instrumental limitations for the ground-state lifetime. An important
characteristic of this configuration is the absence of any ``trap'' state for atomic population. Without the
far-detuned excited state $|\mathrm{e}_{2}\rangle$, the parametric process would be impossible due to trapping
of atoms in state $|-\rangle$  due to optical pumping. The presence of the fourth level allows the ``recycling''
of population. In addition, the asymmetry of the two transitions in terms of Rabi frequency and detuning results
in different AC Stark shifts of the ground states, potentially allowing for the efficient correlation between
noise sidebands at low frequencies.

A few recent experiments reported observation of VS via PSR squeezing on different transitions in
Rb~\cite{ries_experimental_2003,mikhailov2008ol}. However, the amount of observed squeezing was significantly
smaller than predicted, and some experimental efforts failed to observe any squeezing at
all~\cite{zibrovJETP2004,hsu_effect_2006}. All these measurements were degraded by excess noise caused by
interaction of light and atoms, which was not accounted for in the original theory. In this manuscript we present
experimental and theoretical study of the noise quadratures of light in PSR squeezing configuration after its
propagation through the cell, and analyze their dependencies on atomic density, laser power and detuning.

%\section{Experimental Setup}

\textbf{Experimental Setup.} The schematic of the experiment is shown in Fig.~\ref{fig:apparatus}(a). A
commercial Vortex external cavity diode laser ($\approx 7$~mW total power) served as a master laser for
injection-locked high-power slave laser diode, increasing laser power at the cell up to $\approx 40$~mW. No
additional noise associated with the locking process was observed. The laser was tuned across the Rb $D_1$ line
$5^2S_{1/2}\rightarrow5^2P_{1/2}$ ($\lambda \simeq 795$~nm). Before entering the cell the laser beam passed
through a high quality Glan polarizer (GP) to purify its linear polarization. A half-wave plate ($\lambda/2$)
placed before the input polarization beam splitter allowed for smooth adjustments of the pump field intensity.
A pair of lenses $\mathrm{L}$ focused the laser beam inside a cylindrical Pyrex cell with the estimated
minimum laser beam diameter inside the cell of $0.175\pm0.015$~mm FWHM. The cell contained isotopically enriched
$^{87}$Rb vapor, its length and diameter were correspondingly 75~mm and 22~mm, and its windows were tilted at
about 10$^\circ$ to prevent backward reflections.
%Most of the
%presented data were obtained using a cell containing only Rb vapor and no buffer gas (a vacuum cell), but the
%cell with 5~Torr of Ne buffer gas has been also tested.
% (TT-RB87/Ne-2.5T-75-Q-AR cell from Triad Technology Inc.).
The cell was mounted inside a three-layer magnetic shielding to minimize stray magnetic fields, and
the Rb vapor density was controlled by changing the temperature of the cell using a resistive heater
wrapped around the innermost magnetic shield.

After the cell the electromagnetic field in orthogonal polarization (hereafter designated as vacuum field) was
separated on a polarizing beam splitter (PBS), and its noise properties were analyzed using homodyne
detection. The transmitted pump field played the role of a local
oscillator (LO), which was attenuated and
brought to the same polarization as the vacuum field using  another Glan polarizer and a half wave plate combo.
We then mixed these two fields at a 50/50 non-polarizing beam
splitter(NPBS), and directed the resulting two beams to a
home-made balanced photodetector (BPD) with  a transimpedance gain of $10^4$~V/A, 1~MHz 3~dB bandwidth and
electronic noise floor located at 6~dB below shot noise at low
frequencies. The BPD incorporated two matched Hamamatsu S5106
photodiodes with quantum efficiency $\eta=87$\%  and a low noise high bandwidth TI OPA842 operational amplifier.

%\section{Experimental investigation of vacuum squeezing}

\textbf{Experimental investigation of vacuum squeezing.} Fig.~\ref{fig:apparatus}(c) shows typical dependence of
the detected noise power on the LO relative phase controlled by a mirror placed on a piezo-ceramic
transducer (PZT). As the LO phase is scanned, the measured noise signal oscillates between minimum and maximum
values corresponding to the detection of the quadrature with reduced or increased noise. One can see that the
minimum value of noise falls below the shot noise limit, indicating
that we have successfully produced squeezed vacuum. The level of shot
noise was determined by blocking the squeezed vacuum input to the NPBS.
However, one can also immediately see that this is not a
minimum-uncertainty state, since the maximum noise power exceeds the shot
noise limit by a significantly larger
amount than the reduction of the minimum noise below this limit.

\begin{figure}[h]
  \includegraphics[width=1.0\columnwidth]{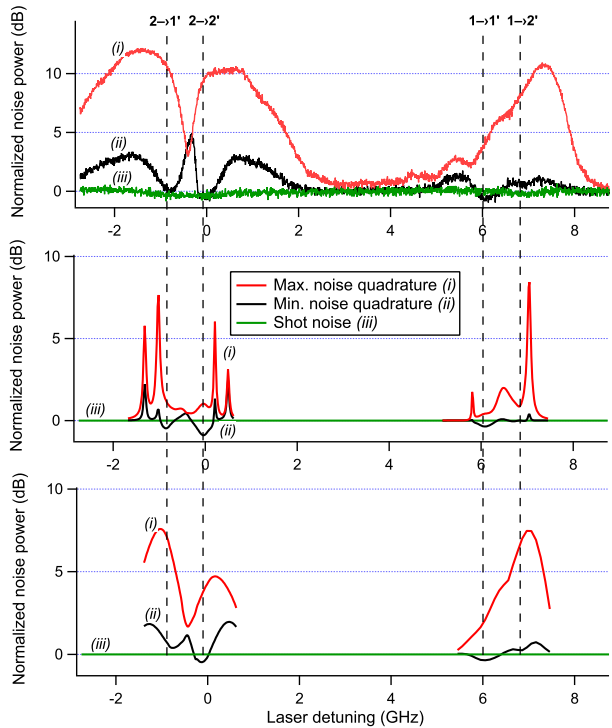}
  \caption{\emph{(a)} Experimental dependence of squeezed and anti-squeezed noise quadratures at 1.2 MHz on
  laser detuning as the laser frequency is scanned across all four transitions of ${}^{87}$Rb
  $\mathrm{D}_1$ line. \emph{(b)} and \emph{(c)} Theoretically calculated noise quadrature
  spectra for noise frequency $\delta = 0.2\Gamma$ with and without Doppler averaging, correspondingly. For
  stationary atoms (no Doppler broadening, \emph{(b)})we used the following parameters: the Rabi frequency $\Omega_f=30\Gamma$, ground state coherence decay rate
  $\gamma_0=0.01\Gamma$, and the cooperativity parameter $C=100$. Same  $\Omega_f=30\Gamma$ and
  $\gamma_0=0.01\gamma$ were used for moving atoms (\emph{(b)}), although the cooperativity parameters was
  increased to $C=1000$ to take into account distribution of atoms in different velocity groups (see discussion in the text).
    }
  \label{fig:sq_vs_detuning_vac}
\end{figure}

Fig. \ref{fig:sq_vs_detuning_vac}(a) shows both minimum and maximum values of the noise quadrature as
the laser frequency is tuned across all four atomic transitions of the ${}^{87}$Rb D1 line. For each measurement
the phase of the local oscillator was locked to the corresponding extremum at $1.2$~MHz by
noise-locking technique~\cite{mckenzie_quantum_2005} with 1~kHz dither frequency.

It is easy to see that there is a large difference between the maximum and
minimum noise quadrature values, showing that the detected noise indeed
depends on the phase of the local oscillator.
However, even the minimum quadrature falls below the shot noise limit only
in a few small regions of
laser frequency, and for the rest it exceeds this limit, sometimes by several dB. This excess noise
results from the interaction of pump light with atoms. The existence of this excess noise complicates the observation of squeezing, since now
to suppress the value of noise quadrature below the shot noise limit, the reduction of noise via PSR  must overcome the additional noise due to interaction with atoms. % I am not convinced by this argument. To be discussed (AL).

\begin{figure}[h]
\includegraphics[angle=0,width=1.0\columnwidth]{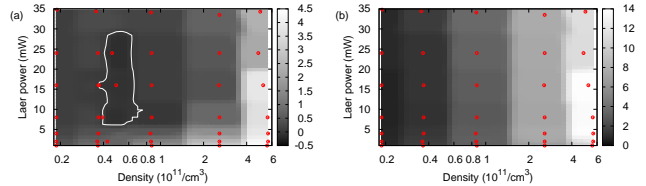}
  \caption{Experimentally measured minimum~\emph{(a)} and maximum~\emph{(b)} values of the
  noise quadrature relative to shot noise (in dB) as functions of laser power and atomic density. The laser frequency is fixed at
  the $F=2\rightarrow F'=2$ transition, and the detection frequency is $1.2$~MHz. Red circles indicates the
  experimental points with gradient extrapolation between these points. The white contour in \emph{(a)} shows the
  parameter range where the minimum quadrature was lower than the shot noise.
    }
  \label{fig:contrast_det2_after_foc}
\end{figure}

To study the balance between the squeezing via PSR and the atomic excess noise we have fixed the laser
near $F=2\rightarrow F'=2$ transition, and then measure the minimum and maximum values of noise
quadrature in a range of laser powers and atomic densities. The resulting dependence is shown in
Fig.~\ref{fig:contrast_det2_after_foc}.

The measurements show that at very low atomic density the difference between the minimum and maximum
values of noise quadrature is small, but it grows monotonically with atomic density. However, the
excess noise grows as well, and as a result squeezing occurs only in the narrow range of fairly low
atomic densities. For higher densities even the minimum noise increases above the shot noise,
reaching up to $4-6$~dB of excess noise. Changes in the laser power produce similar behavior: for
very low pump powers there is no squeezing, probably due to resonant absorption, the noise reaches its
minimum in an intermediate power range, and then it increases again at higher powers.

Fig.~\ref{fig:contrast_det2_after_foc}(a) shows that for our experimental
parameters VS occurs in a very small ``island'' of the parameter space, and only excess noise is measured
elsewhere. This may explain the inconsistency between earlier experimental
works~\cite{ries_experimental_2003,zibrovJETP2004,hsu_effect_2006,mikhailov2008ol}, since it is not
hard to miss the optimal conditions for squeezing detection. We also noticed that the observation is
quite sensitive to various experimental settings: for example, small longitudinal displacement of the
laser focus inside the cell has changed the position of the ``squeezing island'' in the laser
power/atomic density parameter plane.\\

%\subsection{Numerical simulations}

\textbf{Numerical simulations of quantum noise.} The numerical simulations presented below are a direct
extension of the procedure described in \cite{lezama_numerical_2008}.
This model was developed for the
calculation of the quadrature dependent noise power spectrum of two orthogonal polarizations of light traversing
an atomic medium taking into account the complete Zeeman-sublevel structure of a given two-level atomic
transition. The atomic medium is characterized by the cooperativity parameter $C=\frac{N\eta^2L}{c\Gamma }$
where $N$ is the number of atoms, $L$ the sample length, $c$ the speed of light in vacuum, $\Gamma$ the excited
state decay rate and $\eta$ the characteristic atom-field coupling constant (half the single photon reduced Rabi
frequency).
In this calculation, the evolution of the operators describing the field and the atomic medium is described
through Heisenberg-Langevin equations~\cite{CT_book} thus explicitly taking into account quantum fluctuations of
the atomic variables which are the source of excess noise~\cite{hsu_effect_2006,lezama_numerical_2008}.

%Two contributions are considered for the light fluctuation evolution. The
%first, already considered in \cite{matsko_vacuum_2002} is the semiclassical effect of the mean atomic response
%on the evolution of the field fluctuations. This semiclassical contribution  corresponds to the PSR and is the
%source of vacuum squeezing. In addition to the semiclassical contribution, our calculation explicitly
%incorporates the effect of the quantum fluctuations of the atomic operators via appropriate Langevin force
%terms. The atomic fluctuation contributions are detrimental to squeezing and can result in significant excess
%noise.
%Is there insufficient room for this paragraph?  I found it useful. (TN)

Previous work~\cite{lezama_numerical_2008} focused on the quantum noise of light interacting with an individual
hyperfine transition. Although spectral features corresponding to VS were predicted for some transitions, the
results of these simulations differ significantly from the experimental observations reported here. A clear
example of this difference is given by the fact such calculations predict no squeezing at the $F=1
\rightarrow F'=1$ transition where squeezing is observed to be present (See Fig.~\ref{fig:sq_vs_detuning_vac}).
The origin of this discrepancy is in the oversimplification of the atomic level structure. We show in the
following that the complete excited state hyperfine structure is essential for an accurate quantitative
description of the light-atom interaction~\cite{rochester_self-rotation_2001}. In consequence,  we have extended
the previous calculations~\cite{lezama_numerical_2008} to include a realistic description of the $^{87}$Rb D1
line by including both transitions from a given ground state hyperfine level to the two excited state hyperfine
levels (including the complete Zeeman substructure). The amplitude of the field is given by the reduced Rabi
frequency $\Omega_f$ of the fine transition $5S_{1/2} \rightarrow 5P_{1/2}$.

\begin{figure}[h]
\begin{center}
\includegraphics[angle=0,width=1\columnwidth]{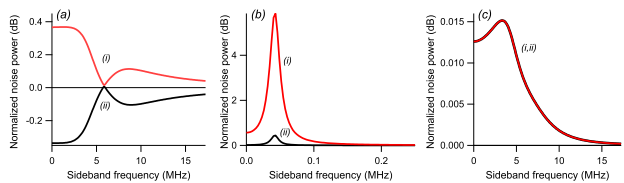}
  \caption{Maximum \emph{(i)} and minimum \emph{(ii)} noise quadrature for the laser tuned to the
  $F=1 \rightarrow F'=1$ transition, if \emph{(a)} complete excited state hyperfine structure is included into the calculations,
  and individual contributions of the $F=1 \rightarrow F'=2$\emph{(b)} and $F=1 \rightarrow F'=1$\emph{(c)}
   transitions alone. ($C=100$, $\Omega_f=10\Gamma$, $\gamma=0.001\Gamma$, $\Gamma=2\pi \times 6$MHz.)
  }
  \label{allsingle}
  \end{center}
\end{figure}

The key role played by off resonant excited state hyperfine transitions is clearly illustrated in
Fig.~\ref{allsingle} where both noise quadratures are shown for a pump field tuned at the $F=1 \rightarrow F'=1$
transition. No squeezing is predicted when each individual transition is
considered separately. However, a significant amount of squeezing, in
agreement with our experimental observations,
is predicted at low noise frequencies when both transitions are
simultaneously taken into account.

The calculated noise power at noise frequency $\delta = 0.2\Gamma$ ($\thickapprox$ 1.2 MHz) on both the squeezed
and anti-squeezed quadratures of the transmitted vacuum field as a
function of laser detuning is presented in
Fig. \ref{fig:sq_vs_detuning_vac}(b) for a homogeneous ensemble of atoms at rest (cold atoms). Squeezing is
present around the $F=2\rightarrow F'=1$, $F=2\rightarrow F'=2$ and $F=1\rightarrow F'=1$ transitions. Almost no
squeezing is observed around the $F=1\rightarrow F'=2$ transition. An interesting feature of this plot concerns
the sharp structures observed both on the squeezed and anti-squeezed noise spectra. As the optical detuning of
the laser field is changed, the differential AC Stark shifts of the ground state Zeeman sublevels varies. For
specific detunings and light intensities, the energy difference between AC Stark shifted sublevels is brought
into resonance with the observed noise frequency giving rise to significant enhancement of the squeezing and
excess noise. The width of the narrow structures is monotonically dependent on  the ground state decoherence
rate.

As observed in Fig. \ref{fig:sq_vs_detuning_vac}(b), even atoms largely detuned from the optical resonance can
significantly contribute to the squeezing and excess noise spectra. In consequence, the accurate comparison with
experimental results obtained in a vacuum cell, necessarily requires the consideration of the full atomic
velocity distribution. We have numerically integrated the velocity dependent contributions of all atoms in the
Doppler profile following the procedure described in \cite{lezama_numerical_2008}. The results are presented in
Fig. \ref{fig:sq_vs_detuning_vac}c. For this calculation we have used the same parameters as for Fig.
\ref{fig:sq_vs_detuning_vac}b except for the cooperativity parameter that was increased to $C=1000$ in order to
maintain a significant number of atoms for each velocity class. As expected, the spectral features are smoothed
and broadened and a significant reduction of the squeezing is observed in comparison with Fig.
\ref{fig:sq_vs_detuning_vac}b. Considering that no fine tuning of the parameters was intended, the overall
agreement between the experimental observation and the numerical simulation is quite satisfactory.

\begin{figure}[h]
\begin{center}
\includegraphics[angle=0,width=1.0\columnwidth]{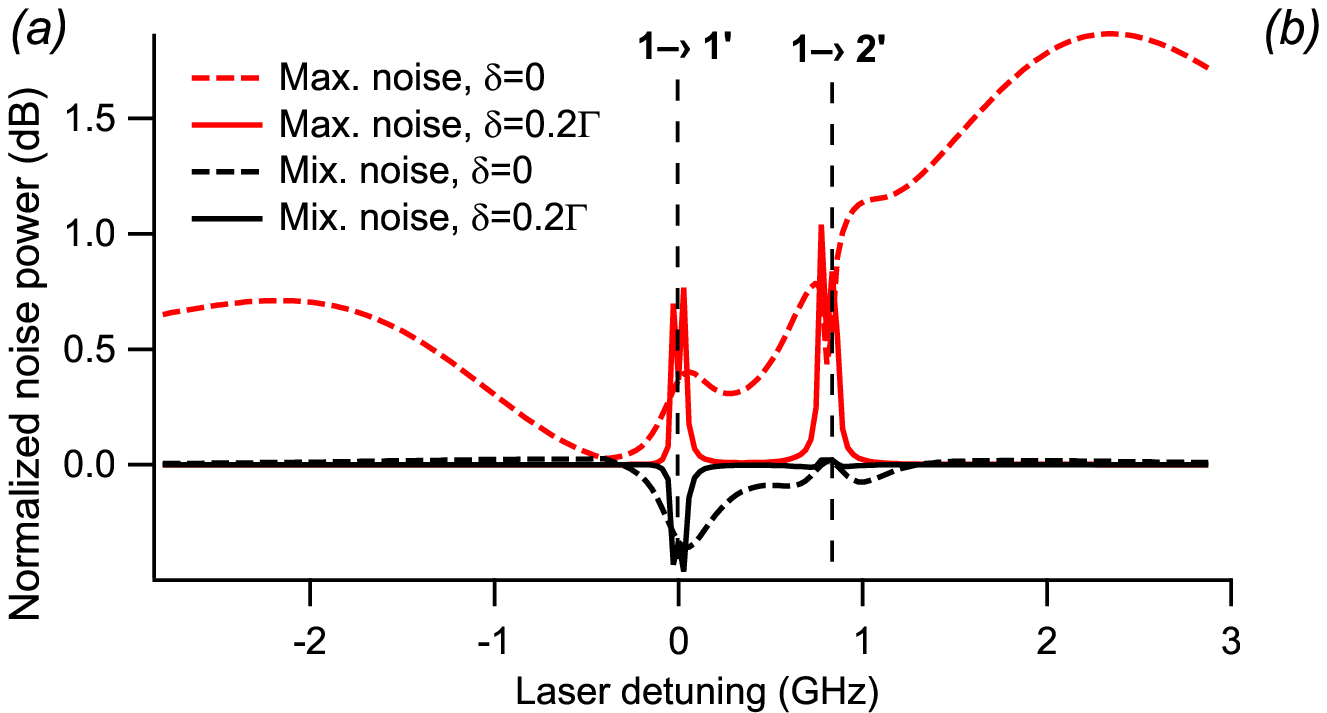}
\includegraphics[angle=0,width=1.0\columnwidth]{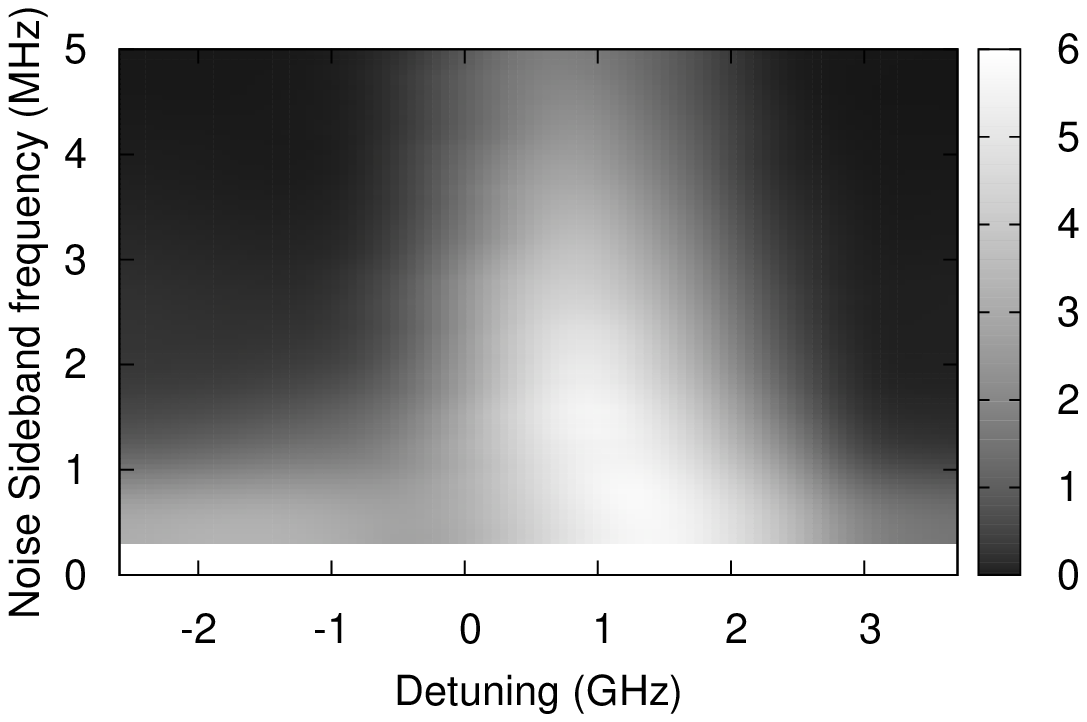}
 \caption{\emph{(a)} Calculated minimum and maximum noise power
  as a function of the laser detuning for noise sideband frequencies $\delta =0$
 and $\delta = 0.2\Gamma$ ($C=100$, $\Omega_f=10\Gamma$, $\gamma=0.001\Gamma$, no Doppler average). \emph{(b)}
 Experimentally measured maximum noise of the anti-squeezed quadrature
relative to shot noise (in dB) as a function of the sideband frequency and laser detuning. }
  \label{lowfreq}
\end{center}
\end{figure}

The contribution of far detuned atoms to the vacuum fluctuations depending on the noise frequency $\delta$ is
illustrated in Fig.~\ref{lowfreq}(a) where the squeezed and anti-squeezed quadratures are calculated for atoms
at rest ($C=100$, $\Omega_f=10\Gamma$, $\gamma=0.001\Gamma$) for two different noise frequencies $\delta = 0$
and $\delta = 0.2\Gamma$. Notice the large contribution to the noise at zero frequency arising from atoms largely
off resonance. We interpret the broad range of excess noise to resonant, nearly elastic scattering of photons
from the pump field polarization into the orthogonal polarization. The experimentally measured noise spectrum
plotted in Fig.~\ref{lowfreq}(b) shows a similar trend.

%\section{Outlook and conclusions}

\begin{figure}[h]
  \includegraphics[angle=0,width=1.0\columnwidth]{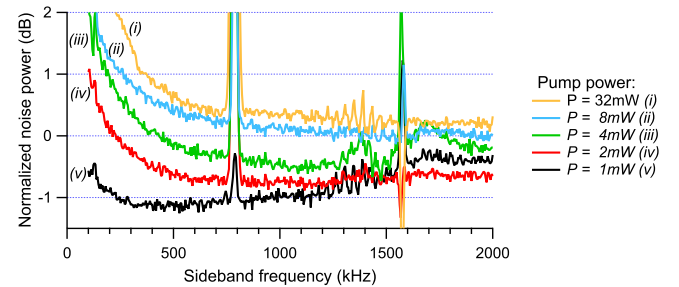}
  \caption{Minimum quadrature noise vs sideband frequency for different pump power
    in the vapor cell containing ${}^{87}$Rb and $5$~Torr of Ne buffer gas. Shot noise corresponds to 0~dB.
    Laser frequency is set to the
    optimum squeezing point near $F=1 \to F'=2$, and the temperature of the cell is $69^\circ$C
    ($N=5.1\cdot10^{11}~\mathrm{cm}^{-3}$).
    Extra noise peaks at 700 and 1600 kHz  are due to environmental noise.
    }
  \label{fig:record_squeezing}
\end{figure}

\textbf{Outlook and conclusions.} Both theory and experiment agree that
the excess noise resulting from
light-atom interaction seriously limits the overall quantum noise suppression. Thus, to improve the performance
of the PSR squeezing, it is crucial to identify ways to minimize the excess noise. For example, our
numerical simulations with no Doppler broadening predict that vacuum squeezing can be improved by switching to
cold atoms (see Fig.~\ref{fig:sq_vs_detuning_vac}). Another option for improvement is to reduce the ground state
relaxation rate by adding a small amount of a buffer gas to the Rb cell. While the amount of self-rotation does
not depend on ground-state coherence ~\cite{matsko_vacuum_2002}, smaller absorption should improve the observed
squeezing. The highest squeezing to date was measured in a cell with
$5$~Torr of Ne buffer gas with the laser
tuned near the $F=2\rightarrow F'=1$ transition (see Figure
\ref{fig:record_squeezing}).
While there is a clear improvement in the squeezing performance when the buffer gas pressure is relatively low
($\le 5$~Torr), there should exist an optimal pressure. In the high-pressure limit the interaction of atoms with
both excited states reduces the transparency, and can even lead to enhanced absorption ~\cite{novikovaJOSAB05}.

In conclusion we present the experimental and theoretical analysis on noise characteristics of vacuum
field propagating through the resonant atomic vapor under PSR conditions. We have found that even
though it is possible to reduce the minimum quadrature noise below the standard quantum limit, the
 interaction of pump field with atoms adds some extra noise to the vacuum field, and makes the
 observation of squeezing possible only in a very narrow parameter space. The experimental
 observations are in very good agreement with the numerical simulations,
in which the exact energy
 level structure and the Doppler effect due to atomic thermal motion were taken into account.

This research was supported by NSF grants PHY-0758010 and PHY-0755262 (REU), Jeffress Research grant J-847, the
College of William~\&~Mary, CSIC, PEDECIBA, and Fondo Clemente Estable.
% (Uruguayan agencies).

%%%%%%%%%%%%%%%%%%%%%%%
%\vspace{12pt}

%\bibliography{bibliography}
%\bibliography{psrsq}
%\bibliographystyle{tMOP}
%\vspace{12pt}

%%%%%%%%%%%%%%%%%%%%%%%

\end{document}